\documentclass[twoside,twocolumn,superscriptaddress,english,pra,showpacs]{revtex4-1}
\usepackage[T1]{fontenc}
\usepackage[latin9]{inputenc}
\usepackage{amssymb}
\usepackage{graphicx}
\usepackage{color}
\usepackage[normalem]{ulem}
\usepackage{dcolumn}
\usepackage{bm}
\usepackage{natbib}
\usepackage{amsmath}
\usepackage{appendix}
\usepackage{here}

\makeatletter

 \@ifundefined{textcolor}{}
 {%
   \definecolor{BLACK}{gray}{0}
   \definecolor{WHITE}{gray}{1}
   \definecolor{RED}{rgb}{1,0,0}
   \definecolor{GREEN}{rgb}{0,1,0}
   \definecolor{BLUE}{rgb}{0,0,1}
   \definecolor{CYAN}{cmyk}{1,0,0,0}
   \definecolor{MAGENTA}{cmyk}{0,1,0,0}
   \definecolor{YELLOW}{cmyk}{0,0,1,0}
 }

\makeatother

\usepackage{babel}
\begin{document}

\title{
Quantum annealing via  environment-mediated quantum diffusion}

\author{Vadim N. Smelyanskiy}
\affiliation{Google, Venice, CA 90291}
\author{Davide Venturelli}
\affiliation{USRA Research Institute for Advanced Computer Science (RIACS),  Mountain View CA 94043}
\affiliation{NASA Ames Research Center, Mail Stop 269-1, Moffett Field CA 94035-1000}
\author{ Alejandro Perdomo-Ortiz}
\affiliation{University of California Santa Cruz, University Affiliated Research Center at NASA Ames}
\affiliation{NASA Ames Research Center, Mail Stop 269-1, Moffett Field CA 94035-1000}
\author{ Sergey Knysh}
\affiliation{Stinger Ghaffarian Technologies Inc., 7701 Greenbelt Rd., Suite 400, Greenbelt, MD 20770}
\affiliation{NASA Ames Research Center, Mail Stop 269-1, Moffett Field CA 94035-1000}
\author{ Mark I. Dykman}
\affiliation{Department of Physics and Astronomy, Michigan State University, East Lansing, MI 48824-232}

\date{\today}
\begin{abstract}

 We show that quantum diffusion near the quantum critical point can provide a highly very efficient mechanism of open-system quantum annealing.  It is based on the diffusion-mediated recombination of excitations.   For an Ising spin chain coupled to a bosonic bath, excitation diffusion in a transverse field sharply  slows down  as the system moves away from the quantum critical region. This leads to spatial correlations and effective freezing of the excitation density. We find that obtaining an approximate solution via the diffusion-mediated quantum annealing can be faster than via closed-system quantum annealing or Glauber dynamics.

\end{abstract}
\maketitle

Quantum annealing (QA)   has been proposed as   a candidate for a speedup of solving hard optimization  problems \cite{Nishimori:98,Brooke:99,Farhi:02}. Optimization can be thought of as motion toward the potential minimum in the energy landscape associated with the computational problem.  Conventionally, QA is related to quantum   tunneling in the landscape that is slowly varied in time \cite{QA-book}. It provides an alternative to simulated annealing, which relies on classical diffusion via thermally activated interwell transitions. 
 It was suggested  that  the coupling to the environment would not be  necessarily detrimental to QA \cite{Averin:09,Dwave:11,DWaveNature:2013} .
 
Recently the role of quantum tunneling as a computational resource has become a matter of active debate  \cite{Santoro:02, Troyer:2015, Nishimori:2008, Hastings:2013, Crosson:2014}, as it is not necessarily advantageous compared to  classical computational techniques, e.g., the path integral Monte Carlo \cite{ Issakov:2015}.  In addition, dissipation and noise can make tunneling incoherent, significantly slowing down  \cite{Kagan_Leggett92} the transition rates that underlie QA.

In this paper we show that dissipation-mediated quantum diffusion 
can provide an efficient alternative resource for QA. We model QA as the evolution of a multi-spin system with a time-dependent Hamiltonian. The diffusion involves environment-induced transitions between entangled states. These states are  delocalized coherent superpositions of multi-spin configurations separated by a large Hamming  distance. At a late stage of QA the diffusion coefficient decreases. Ultimately diffusion becomes hopping  between localized states  and QA is dramatically slowed down. An important question is whether the solution obtained by then is closer to the optimum than the solution obtained over the same time classically.

Diffusion plays a special role where the system is driven through the quantum critical region, as often considered in  QA \cite{QA-book,Brooke:99,Santoro:02}. A well-known result of going through such a region is generation of excitations via the Kibble-Zurek mechanism \cite{Kibble}. This leads to an error, in terms of  QA, as the system is ultimately frozen in the excited state. The generation rate can be even higher in the  presence of coupling to the environment \cite{Santoro:08}. 

It is diffusion that makes it possible for the excitations to ``meet'' each other and to recombine, thus reducing their number. Near the critical region diffusion is enhanced because of the large correlation length. It has universal features related to the simple form of the excitation energy spectrum. 

The novel effect  of quantum-diffusion induced acceleration of QA is of utmost importance for systems with {\it delocalized} multi-spin excitations. To reveal and characterize this effect, we study it  here for a model  with no disorder.
The specific model is a one-dimensional Ising spin chain, where the spins are coupled to the environment and the system is driven through the quantum phase transition by varying a transverse magnetic field. Among recent applications of this classic model we would mention cold atom systems \cite{Orth:08,Cirac:13,Saffman:13}  and the circuit QED \cite{Marquardt13}. 

We assume that each spin is weakly coupled to its own bosonic bath. The QA Hamiltonian is
\begin{equation}
\label{eq:QA}
H_{QA}=-J\sum_{n=1}^{N-1}(\sigma_{n}^{z}\sigma_{n+1}^{z}+g\sigma^{x}_{n})-\sum_{n=1}^{N}\sigma^{x}_{n} X_n+H_B,
\end{equation}
\noindent
where $N$ is the number of spins, $J g(t)$ is the transverse field,  $\sigma^{x}_{n},\sigma^{z}_{n}$ are Pauli matrices, $H_B=\sum_{n,\gamma}\hbar\omega_{\gamma n} b_{\gamma n}^{\dagger}b_{\gamma n}$ is  the baths Hamiltonian; $X_n=\sum_{\gamma}\lambda_{\gamma n}(b_{\gamma n}^{\dagger}+b_{\gamma n})$,
and $b_{\gamma n}^{\dagger},b_{\gamma n}$ are boson creation/annihilation operators  in  the $n$th   bath.
We assume Ohmic dissipation,  2$\sum_{\gamma}(\lambda_{\gamma n}/\hbar)^{2}\delta(\omega-\omega_{\gamma n})$=$\alpha\omega$, $\alpha\ll 1$,
and  linear schedule for reducing the transverse field, $\dot g(t) $=$ -v$<0,  starting from  the initial value $g_i\gg 1$. We further assume translational symmetry, so that $\lambda_{\gamma n}, \omega_{\gamma n}$ are independent of $n$. 
The  spin-boson coupling (\ref{eq:QA}) provides a microscopic model for the classical spin-flip process in the Glauber dynamics \cite{Glauber:63}.

 In the absence of coupling to the environment, model (\ref{eq:QA}) describes  a quantum phase transition between   a paramagnetic phase ($g >1$) and a ferromagnetic phase ($g< 1$) \cite{Sachdev:99}.
The spin part of the  Hamiltonian (\ref{eq:QA}) can be  mapped onto  fermions
\cite{Lieb:61} using the Jordan-Wigner transformation, $\sigma_{n}^{x}=1-2a_{n}^{\dagger}a_n$, $\sigma_{n}^{z}=-K(j)(a_{n}^{\dagger}+a_{n})$ where $K(j)=\prod_{i<j}\sigma_{i}^{x}$;   $a_{n}^{\dagger}$ and $a_n$ are fermion creation and annihilation operators. Changing in the standard way to
new creation and annihilation operators
$\eta_{k}^{\dagger}$, $\eta_{k}$, with $\eta_k=\frac{1}{\sqrt{N}}\sum_{n=1}^{N}[a_n\cos(\theta_k/2)-i a_{n}^{\dagger}\sin(\theta_k/2)]e^{-i k n}$, we obtain the Hamiltonian of the isolated spin chain as  $H_0=2J \sum_{k}\epsilon_k\eta^{\dagger}_{k}\eta_{k}$, where $\epsilon_k$ is the dispersion law in the fermion band,
\begin{equation}
 \epsilon_k=\sqrt{(g-\cos k)^2+\sin^2 k},\quad \tan\theta_k=\frac{\sin k}{g-\cos k}.\label{eq:ek}
\end{equation}

In the course of QA,  pairs of fermions with opposite momenta are born from vacuum due to the  Landau-Zener transitions as the system passes through  the critical point $g=1$ \cite{Kibble}.  The resulting density of  excitations  $n_v$ in the thermodynamic limit is simply related to the QA rate $v$ \cite{Jacek:2005},
\begin{equation}
n_v=  (\hbar v/8\pi J)^{1/2}.\label{eq:KZM}
\end{equation}

Coupling to  bosons leads to relaxation of the fermion system and renormalization of its spectrum. From Eq.~(\ref{eq:QA}), the coupling Hamiltonian in terms of the fermion operators has the form
\begin{align}
\label{eq:H_i}
H_i=&\sum_{kk'}h_{kk'}X_{k-k'}, \quad h_{kk'}= c_{kk'}\eta_k^\dagger \eta_{k'}  
\nonumber\\
&+  s_{kk'}\eta_k^\dagger\eta_{-k'}^\dagger + s^*_{k'k}\eta_{-k}\eta_{k'},
\end{align}
where $X_q=\sum_\gamma \lambda_\gamma( b_{\gamma q} + b_{\gamma\, -q}^\dagger)$ are boson field operators, $  b_{\gamma q}=N^{-1/2}\sum_{n}b_{\gamma n}\exp(-iqn)$; the  coefficients $c_{kk'}$ and $s_{kk'}$ are expressed in terms of the rotation angles $\theta_k, \theta_q$, see Eq.~(27) of the Supplemental Material (SM).

From Eq.~(\ref{eq:H_i}) one can identify two types of relaxation processes. The first is intraband scattering in which a fermion momentum is transferred to bosons. The rate of intraband scattering $k\to k'$ is $W^{+-}_{k k'} \propto |c_{k k'}|^2$. The second process is interband transitions of generation and recombination of pairs of fermions with rates $W^{++}_{k k'}$ and $W^{--}_{k k'}$, respectively; both rates are $\propto |s_{k k'}|^2$,
\begin{align}
W_{k k'}^{\mu\nu} &=(2\pi \alpha/N)\Omega_{k k'}^{\mu\nu} \left[1-\mu\nu \cos(\mu\theta_k - \nu\theta_{k'})\right]\nonumber\\
&\times[\bar n(\Omega_{k k'}^{\mu\nu})+1],\quad
\Omega_{k k'}^{\mu\nu} =
2J(\mu \epsilon_k+ \nu \epsilon_{k'})/\hbar,
\label{eq:W}
\end{align}
\noindent
where $\mu,\nu$=$\pm $ and $\bar n(\omega)=[\exp(\hbar\omega/k_B T)-1]^{-1}$.

The single-particle  quantum kinetic equation that incorporated these processes was derived in Ref.~\cite{Santoro:08}.  The equation was written for the coupled fermion populations $P_k=\langle \eta^\dagger_k \eta_k\rangle $  and coherences $\langle \eta_k \eta_{-k}\rangle$.  It  involved two major approximations, the spatial uniformity of the fermion distribution and the absence of fermion correlations. These approximations hold in the critical region, where the gap in the energy spectrum $\Delta(g)=2J|1-g| < k_BT$.
For a sufficiently low QA rate, the density of excitations is dominated by thermal processes rather than the Landau-Zener tunneling \cite{Santoro:08}. The fermion population is $ P_k=1/[\exp(2J \epsilon_k/k_B T)+1]$.

The goal of QA is to reduce the number of excitations, which happens after the system goes through the critical region. As we show, a significant reduction can be achieved already very close to the critical region. However, the approximation \cite{Santoro:08} does  not describe the dynamics in this range where many-fermion effects and the associated spatial density fluctuations become significant. These effects can be described by the Bogoliubov hierarchy  of equations for many-particle Green's functions. However, as we show,  the relevant for the QA scaling relations between the speed $\dot g$ and the final density of excitations can be found in a simpler way.

Behind the critical region, where 
\begin{equation}
e^{- \Delta(g)/k_B T}\ll 1,\quad \Delta(g)=2J|1-g|,
\label{eq:semi}
\end{equation}
the fermion density $n\equiv \langle n(x)\rangle $ becomes small and the average inter-fermion distance largely exceeds the thermal wavelength  $\lambda_T$=$2\pi \hbar/(2m_e k_BT)^{1/2}$ [$m_e$=$ \hbar^2(1-g)/2Jg$ is the effective mass]. As the first approximation,  one can describe the  fermion system by the single-particle Wigner probability density
$\rho_W(x,k)=(2\pi)^{-1}\int dp\langle \eta^\dagger_{k+p/2}\eta_{k-p/2}\rangle e^{-ipx}.
$
For weak coupling to the bosonic bath, the kinetic equation for this function  reads,
\begin{align}
\label{eq:QKE}
\partial_t \rho_W +(\partial_k\epsilon_k)\partial_x\rho_W = \hat{\cal L}^{(0)}\rho_W + \hat{\cal L}^{(1)}\rho_W, 
\end{align}
where operator $\hat{\cal L}^{(0)}$ describes intraband scattering,
\begin{align}
\label{eq:intraband}
\hat{\cal L}^{(0)} \rho_W(x,k)\approx &\frac{N}{2\pi}\int dq \left[W^{+-}_{qk}\rho_W(x,q)
-W^{+-}_{kq}\rho_W(x,k)\right],
\end{align}
whereas $ \hat{\cal L}^{(1)}$ describes interband transitions  and is discussed below.

The coefficients in Eq.~(\ref{eq:intraband}) simplify close to the critical point, where  $1-g\ll 1$ but $\Delta(g)\gg k_BT$. Here $W^{+-}_{kq}\sim \tau_r^{-1}$, where $\tau_r^{-1}$ is the momentum relaxation rate,
\begin{align}
\label{eq:relaxation_rate}
\tau_{r}^{-1} = 2\alpha k_BT[(1-g)/\beta g \hbar^2]^{1/2}, \quad \beta=2J/k_BT.
 \end{align}
Using the explicit form of $W^{+-}_{kq}$, one can show that the eigenvalues of the operator $\hat {\cal L}^{(0)}$ are non-positive. The zero eigenvalue corresponds to the Maxwell distribution over momentum, $\rho_W(x,k)\propto\exp(-\beta\epsilon_k)$, whereas the next eigenvalue is negative and is separated by a gap $\approx -6.6\tau_r^{-1}$. 

The rate $\tau_r^{-1}(g)$ increases with the distance $1-g\propto\Delta$ from the critical point. Extrapolating it back to the critical region $\Delta\simeq k_B T$ we recover the critical scaling 
$(\tau_r^{-1})_{\rm cr}  \simeq 4 J\alpha/\hbar\beta^2$
%
\cite{Santoro:08}. The system passes the critical region isothermally for $J|\dot g|\ll (\tau_r^{-1})_{\rm cr} $. 

On the time scale large compared to $\tau_r$ the distribution $\rho_W$ takes a simple form of a product of the Maxwell-Boltzmann distribution over kinetic energy and, generally, a coordinate-dependent density $n(x,t)$,  $\rho_W=n(x,t)\exp(-\beta\epsilon_k)/\sum_k\exp(-\beta\epsilon_k)$. If we disregard the term $\hat{\cal L}^{(1)}$ in Eq.~(\ref{eq:QKE}), we obtain a standard diffusion equation for the fermion density
\begin{align}
\label{eq:diffusion}
\dot n(x,t)  =D\partial_x^2 n(x,t), \quad D=c_D 
\frac{J \beta^{1/2} }{\alpha\hbar }\frac{g^{3/2}}{(1-g)^{3/2}},
\end{align}
where $c_D\approx 0.17$ (see Sec.~I in SM for details).  We note that the diffusion   coefficient $D\sim k_B T\tau_r/m_e$ sharply increases near the critical point. 

We now discuss the term $\hat{\cal L}^{(1)}\rho_W$ that describes interband transitions in Eq.~(\ref{eq:QKE}). In the adopted approximation where we disregard fermion correlations and decouple many-fermion Green's functions, we can write this term as a sum of the generation and recombination terms.
The generation term  $[\hat{\cal L}^{(1)}\rho_W(x,k)]_{\rm gen}$ is proportional to the coefficients $W^{--}_{kq}\propto \exp(-\Delta(g)/k_BT)$. It rapidly falls off as the control parameter $g$ moves away from the critical point. Generation of fermions becomes  slow for large $\Delta/k_BT$. The recombination term $[\hat{\cal L}^{(1)}\rho_W(x,k)]_{\rm rec}= -N\sum_qW^{++}_{kq}\rho_W(x,k)\rho_W(x,q)$ also becomes small in this range, because the number of fermions becomes small.

From the above arguments, using the fact that  the distribution over fermion momentum is of the Maxwell-Boltzmann form,  we obtain a standard generation-recombination equation for the spatially-averaged fermion density $\langle n\rangle$,
\begin{align}
\label{eq:generation_recombination}
\langle \dot n\rangle = -w(\langle n\rangle ^2 - n^2_{\rm th})
\end{align}
Here, $n_{\rm th}$ $\equiv$ $ n_{\rm th}(g)$=$N^{-1}\sum_k\exp(-\beta\epsilon_k)$ is thermal equilibrium density, whereas $w(g)=\sum_{k,q}W_{k q}^{++} \exp[-\beta(\epsilon_k+\epsilon_q)]/N n_{\rm th}^2$ is the recombination rate ~\cite{footnote1}. For $\beta\gg 1-g, 1/g$
\begin{equation}
\label{eq:Wbar1}
w(g) \simeq \frac{8\pi\alpha J}{\hbar\,\beta g},\quad
n_{\rm th}(g) \simeq \left(\frac{1-g}{2\pi \beta g}\right)^{1/2}\,e^{-\beta(1-g)}.
\end{equation}

As $g\equiv g(t)$ decreases, the thermal density $n_{\rm th}$ exponentially sharply falls down. The mean density $\langle n\rangle$ cannot follow this decrease, so that the density of fermions becomes higher than the thermal  density. This happens for the value $g(t)=g_0$ where the correction $\delta\langle n\rangle =\langle n(t)\rangle -n_{\rm th}\bigl(g(t)\bigr)$ becomes $\sim n_{\rm th}\bigl(g(t)\bigr)$. The quasistationary solution of the linearized Eq.~(\ref{eq:generation_recombination}) reads $\delta\langle n\rangle \approx -\dot n_{\rm th}/2 w n_{\rm th}$. This gives an equation for $g_0$
\begin{equation}
\label{eq:nonthermal}
\beta^{-1}w(g_0)n_{\rm th}(g_0)=|\dot g|\equiv v.
\end{equation}

For $\exp\{\beta[g_0-g(t)]\} \gg 1 $   we can disregard $n_{\rm th}$ in Eq.~(\ref{eq:generation_recombination}). Then using the explicit form of the rate $w(g)$, we obtain
\begin{equation}
  \label{eq:n}
 \langle n(t)\rangle = \beta^{-1} n_{\rm th}(g_0)/ \log \left[g_0/g(t) \right ].
\end{equation}
Clearly,     $\langle n(t)\rangle $  varies with time only logarithmically.

Another important for the QA consequence of the decrease of $g(t)$ is the sharp decrease of the diffusion coefficient $D=D(g)$, see Eq.~(\ref{eq:diffusion}).  For small $D$, spatial fluctuations of the density $n(x,t)$ become important. They impose a bottleneck on the recombination in one-dimensional systems \cite{Tauber2014}, because for fermions to recombine they first have to come close to each other. In contrast to the usually studied reaction-diffusion systems, in the present case the bottleneck arises not because of the decrease of the density, but, in the first place, because of the falloff of the diffusion coefficient.

Once the recombination becomes  limited by diffusion, the change of the fermion density becomes even slower than in Eq.~(\ref{eq:n}). If we stop decreasing $g$ where thermal generation  can be disregarded, it will take time $\sim N^2/D$ for the density to become  $\lesssim 1/N$. If we then make $g=0$, the system will be in the ground state. Thus the overall time to find a global minimum  of the  optimization problem will be $\propto N^2$. However, this is not our goal.

The density $n_*=\langle n(t_*)\rangle$ where there occurs the crossover to diffusion-limited recombination gives an approximate solution of the QA. It is this solution that we are interested in. As we show, it can be reached in time that is independent of $N$. One can estimate $n_*$ by setting equal the rates $\dot n$ calculated for the recombination and diffusion processes. For the recombination,  one can use Eq.~(\ref{eq:generation_recombination}) written for the local density $n(x,t)$. For the diffusion, one can use Eq.~(\ref{eq:diffusion}) where the mean inter-particle distance $1/\langle n\rangle$ is chosen as a scale on which the density fluctuates. An alternative way of estimating $n_*$ for a time-dependent diffusion coefficient is described in Sec.~IV of the SM. The result reads
\begin{align}
\label{eq:n_star}
n_*= \langle n(t_*)\rangle= kw(g_*))/ D(g_*), \qquad g_*=g(t_*), 
\end{align}
where $k\sim 1$.

Equations (\ref{eq:nonthermal}) - (\ref{eq:n_star}) relate  the crossover value of $g=g_*$ to the value $g_0$ where thermal equilibrium is broken. It is convenient to write them for the scaled distances from the critical value $g=1$, which are given by $x_0=\beta (1-g_0)$ and $x_*=\beta(1-g_*)$,
\begin{align}
\label{eq:x_variables}
&\mu x_0^{1/2} \exp(-x_0)= x_*^{3/2}(x_*-x_0),\nonumber\\
&\mu= c_D\beta^2/8\alpha^2 k\sqrt{2\pi^3},\qquad \mu\gg 1.
\end{align}

Equations (\ref{eq:nonthermal}), (\ref{eq:n}), and (\ref{eq:x_variables}) express the crossover density $n_*$ in terms of the speed of the change of the control parameter $v=|\dot g|$.  Unexpectedly, the dependence of $n_*$ on $v$ is nonmonotonic, see Fig.~1. Our goal is to minimize $n_*$. The optimal value $n_{\rm opt}=\min n_*$ is

\begin{align}
\label{eq:x_*}
n_{\rm opt}\approx \frac{8\pi k\alpha^2}{c_D\beta^3}x_{\rm opt}^{3/2}, \qquad
x_{\rm opt}\approx \ln\mu +1 - \ln(\ln\mu),
\end{align}
where $x_{\rm opt}$ is the value of $x_*$ where $n_*$ is minimal; $x_0=x_{\rm opt}-1$. The optimal speed $v_{\rm opt}$ is related to this value of $x_0$ by Eq.~(\ref{eq:nonthermal}),
\begin{equation}
\label{eq:v_opt}
v_{\rm opt}\approx (64 k\pi^2 J\alpha^3/c_D\beta^5\hbar)\ln(\beta^2/\alpha^2)^{1/2}.
\end{equation}

The above analysis applies for $\beta\gg x_*$ and $x_*-x_0\gg 1$. Therefore the optimal speed of the algorithm is somewhat smaller than $v_{\rm opt}$, and the value of $n_*$ that can be reached is slightly higher than $n_{\rm opt}$, see Fig.~1. However, Eq.~(\ref{eq:x_*}) gives the characteristic scaling of the minimal $n_*$ and the optimal velocity with the parameters. It is seen that $n_{\rm opt}$ is extremely small for weak coupling, $\alpha\ll 1$, and low temperatures, $\beta\gg 1$, and it rapidly decreases with decreasing $\alpha$ and $k_BT/J$.
Importantly, the optimal speed $v_{\rm opt}$ is independent of the size of the system.

It is instructive to compare the optimal speed with the speed $v_{\rm KZ}$ that would lead to the same density $n_v= n_{\rm opt}$ due to the Kibble-Zurek mechanism of creation of excitations in the absence of coupling to the environment. From Eqs.~(\ref{eq:KZM}) and (\ref{eq:v_opt}), 
\begin{equation}
\label{eq:ratio_KZ}
v_{\rm opt}/ v_{\rm KZ} \propto (\beta/\alpha)\ln(\beta/\alpha)^{3/2} \gg 1.
\end{equation}
Therefore the time it takes to reach the approximate solution (\ref{eq:x_*})  in a closed quantum system is much larger than in our case.

\begin{figure}
\hspace*{-0.5cm}\includegraphics[width=8.9cm]{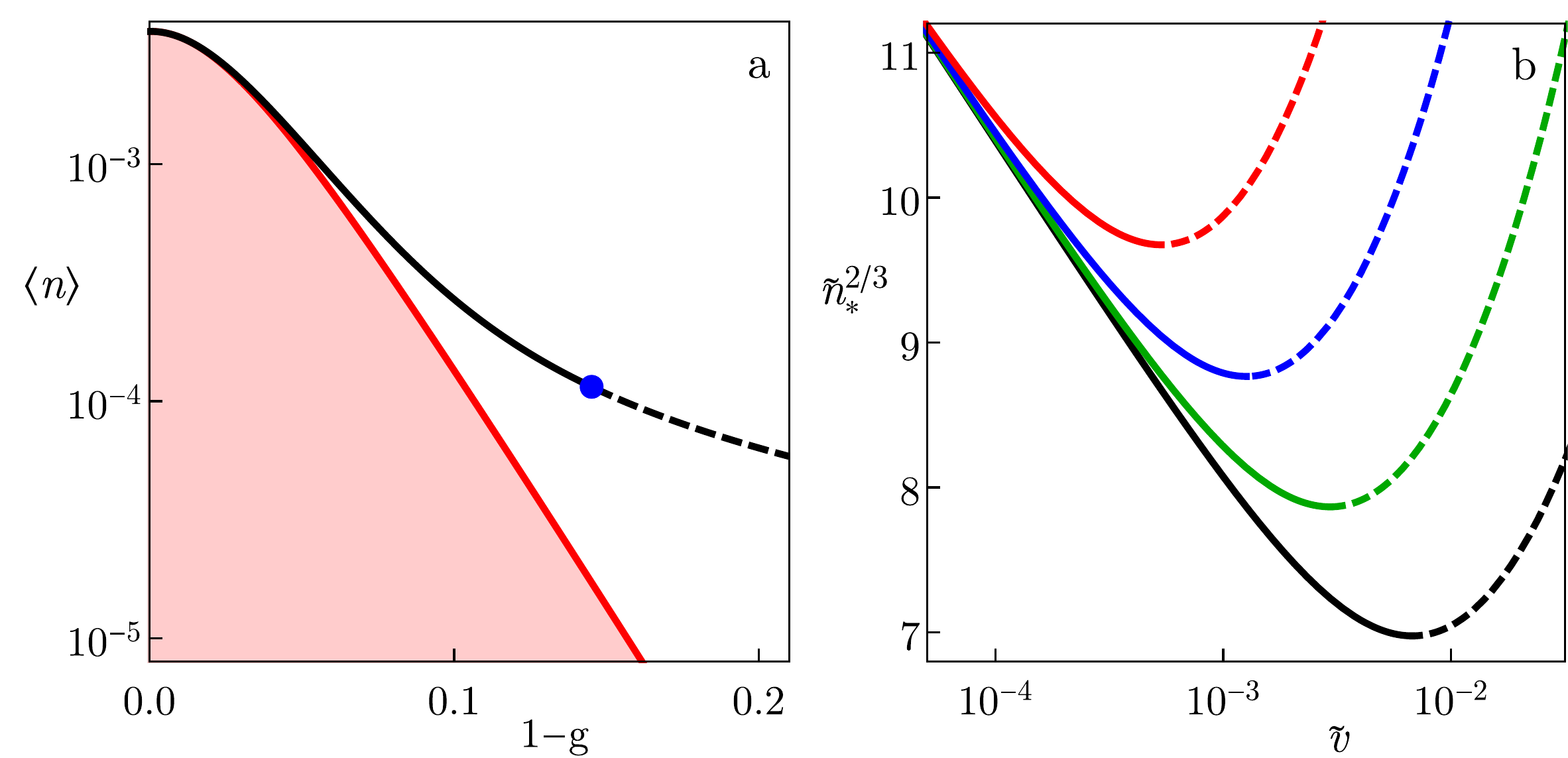}%
\caption{Fermion density vs. the distance to the critical point (a) and vs. the annealing rate (b). In (a), the filled region is bound by the thermal distribution $n_{\rm th}(g)$.  The black line shows the nonequilibrium density $\langle n\rangle$ for $\alpha=0.06$, $\beta=25$ and  $v$=$v_{\rm opt}$=2.85 $\times$ 10$^{-7}$, see Eq.~(\ref{eq:generation_recombination}) \cite{footnote1}.  The blue point marks the crossover value $g_*$. For $g<g_*$ spatial correlations become strong and the theory is inapplicable. 
In (b), the red, blue, green and black  lines show the scaled density $\tilde n_*=c_d\beta^3 n_*/8 k \pi \alpha^2$
vs. the scaled QA rate $\tilde v = \hbar\beta^3 v/4 \sqrt {2\pi} J \alpha$ for  $\log \mu =8,9,10,11$, respectively [parameter $\mu\propto (\beta/\alpha)^2$ is defined in (\ref{eq:x_variables})]. The minimal density $n_{\rm opt}=\min n_*$.  The dashed sections of the lines refer to the regions where the asymptotic theory does not apply.}
\end{figure}

It is instructive also to compare $v_{\rm opt}$ with the speed of annealing based on the classical Glauber dynamics \cite{Glauber:63}. In this dynamics, for low temperatures, $\beta\gg 1$, excitations in the Ising spin chain are eliminated through diffusion of kinks. If the transition rate for a kink to move to a neighboring site is $w_{\rm G}$ and the initial density of the kinks is $\sim 1$, the time $t_{\rm class}$ to reach density $n\ll 1$ is $(8\pi w_{\rm G}n^2)^{-1} $ \cite{Glauber:63}. In terms of our model, the uncertainty relation imposes a limitation $w_{\rm G}\ll J/\hbar$. Therefore the ratio of the times to reach $n_{\rm opt}$ via classical and quantum diffusion is very large, $~\sim t_{\rm class}v_{\rm opt}\propto \beta/\alpha \gg 1$. 

The results demonstrate that quantum diffusion near the critical point provides an important mechanism of the speedup of QA. We find that the bottleneck of QA in an open system can be imposed by the sharp slowing down of the diffusion near  the critical region. The crossover to slow excitation recombination  is accompanied by the onset of significant spatial fluctuations of the excitation density even in the absence of disorder.  At the crossover, the distance to the critical value of the transverse field and the residual density of excitations non-monotonically depend on the quantum annealing rate. Their minimum provides the optimal value of the rate. This value scales with the coupling constant and temperature as  $\alpha^3T^5$, the optimal excitation density is $\propto \alpha^2 T^3$, and the distance to the critical value of the transverse field is $\propto T|\ln T|$. 

For our simple but nontrivial example of QA, attaining the approximate solution  \cite{farhi-app-sol} via the quantum-diffusion mediated process is faster than via  classical diffusion or the closed-system QA.
 One might expect that, in higher-dimensional systems, quantum diffusion over extended states  could provide a route to finding approximate solutions in the presence of disorder.


\hfill

This work was supported in part by the Office of the Director of National Intelligence (ODNI), Intelligence Advanced Research Projects Activity (IARPA), via IAA 145483, by the AFRL Information Directorate under grant F4HBKC4162G001, and by NASA (Sponsor Award Number NNX12AK33A). D.V. and A.P.-O.   were also supported in part by Sandia National Laboratory AQUARIUS project. M.I.D. is grateful to  the NASA Ames Research Center for warm hospitality during his sabbatical and partial support.

\appendix

\begin{widetext}
\vspace{0.4in}
\begin{center}
{\Large\bf Supplemental material}
\end{center}
\end{widetext}
\hfill

\section{Fermion diffusion coefficient}
\label{sec:D} 

In the semiclassical region, Eq.~(6) of the main text, the  fermions have an effective mass $m_e=\hbar^2 |1-g|/2 J g$  and the average thermal velocity $v_T= \sqrt{4g J^2/(1-g)\hbar^2 \beta}$. The  length scale corresponding to a coherent fermion motion is given by  the inelastic mean free path 
 \begin{equation}
  \ell_{\rm mfp}(t)= v_T\tau_r = \frac{ \beta}{\alpha } \,\frac{g(t)}{ 1-g(t) } \gg 1\;, \nonumber
\end{equation}
where $\tau_{r}^{-1}$ is the fermion momentum relaxation rate given in Eq.~(9) of the main text.
 
At sufficiently low density, fermions move mostly independently with typical  thermal wavelength 
 \begin{align}
  \lambda_T=2\pi\hbar /\sqrt{2m_e k_B T}=2\pi  \sqrt{\beta g(t)/2( 1-g(t))}.\label{eq:lT}
  \end{align}
 For small coupling to the bosonic bath and for low temperatures
  \begin{align}
 1 \ll \lambda_T\ll \ell_{\rm mfp} \ll \langle n \rangle ^{-1},
\label{eq:diff-reg}
  \end{align}
  \noindent
where  $\langle n \rangle$  is the mean fermion density.  Two fermions can recombine when  their wave packets  overlap. The recombination probability is $\propto w$, where $w=w(g)$ is given in Eq.~(12) of the main text.

If the  processes of generation and recombination are disregarded, in the parameter range (\ref{eq:diff-reg})  one can describe fermion kinetics using a standard quantum kinetic equation for the single fermion  Wigner probability distribution 
$\rho_W(x,k,t)$=$\frac{1}{2\pi}\int_{-\infty}^{\infty}d\kappa\,    \langle\eta_{k+\kappa/2}^\dagger(t)\eta_{k-\kappa/2}(t)\rangle \,e^{-i \kappa x}$. It reads
\begin{align}
\frac{\partial \rho_W(x,k)}{\partial t}+&\frac{1}{\hbar}\frac{\partial \rho_W(x,k)}{\partial x}\frac{\partial E_k}{\partial k} =\int_{-\infty}^{\infty}dq\,w_{q\rightarrow  k}\,\,\rho_W(x,q) \nonumber \\
&-\rho_W(x,k)\int_{-\infty}^{\infty}dq\,w_{k \rightarrow q}\;,
\label{eq:masterWig}
\end{align}
\begin{equation}
w_{k \rightarrow q}=
\frac{N}{2\pi} W_{k q}^{+-}\;.
\end{equation}
In Eq.~(\ref{eq:masterWig})
\begin{equation}
E_k=2J \epsilon_k\label{eq:Ek}
\end{equation}
\noindent
is the  fermion energy [the scaled energy $\epsilon_k$ is given in Eq.~(2) of the main text]. In the semiclassical region
\begin{equation}
E_k=\Delta + \frac{\hbar^2k^2}{2m_e},\quad \Delta=2J(1-g)\gg k_B T\;,\label{eq:para}
\end{equation}

The stationary solution of  Eq.~(\ref{eq:masterWig}) is given by the spatially-uniform  Boltzmann distribution over the fermion momentum,
\begin{equation}
\rho_W^{(0)} (k)=\frac{1}{Z} e^{-\frac{E_k}{k_B T}}=\frac{1}{Z}e^{-\frac{\Delta}{k_B T}-\frac{K^2}{2}}.\label{eq:PB}
\end{equation}
Here $Z=\int_{-\infty}^{\infty}dk\,\exp(-E_k/k_B T)$,
and we have introduced the scaled momentum $K$,
\begin{equation}
k=   k_{\rm th} \,K ,\quad k_{\rm th}= \left(\frac{1-g}{\beta g}\right)^{1/2}\ll 1\;.\label{eq:K}
\end{equation}

We write the  transition rate  in terms of the rescaled momenta and  expand it in powers of $\beta^{-1}$. To the leading order in $\beta^{-1}$ we have
 \begin{equation}
w_{k\rightarrow q} \approx  w_{k\rightarrow q}^{(0)}=\frac{4 \alpha J}{\hbar\beta} \, \frac{(K^2-Q^2)/2}{1-\exp[-(K^2-Q^2)/2]}\;. \label{eq:wkq}
\end{equation}
In the approximation (\ref{eq:wkq}), the  rate  $w_{k\rightarrow q}$ is symmetric with respect to sign inversion of $k,q$,
\begin{equation}
w_{k\rightarrow q}\approx w_{|k| \rightarrow |q|}^{(0)}\;. \label{eq:sym}
\end{equation}

An important physical argument is that the time evolution of the fermion probability distribution with respect to momentum is fast, it occurs over time $\sim \tau_r$.  The evolution of the spatial distribution (the coordinate-dependent part of $\rho_W$) is much slower. To find this slow evolution, 
we seek the time-dependent solution of Eq.~(\ref{eq:masterWig}) for a weakly spatially nonuniform distrribution $\rho_W(x,k)$ as a sum of symmetric and anti-symmetric  terms with respect to  $k$, with the symmetric part being of the Boltzmann form,
\begin{equation}
\rho_W(x,k,t)=n(x,t)\,\rho_W^{(0)}(k) +\rho_W^{(1)}(x,k,t)\label{eq:expan}
\end{equation}
\noindent
Here
\begin{equation}
n(x,t)=\int_{-\infty}^{\infty} dk \,\rho_W(x,k,t)\;,\label{eq:cN}
\end{equation}
is the spatial probability density
and $\rho_W^{(1)}(x,k,t))=-\rho_W^{(1)}(x,k,t))$ is a term that corresponds to a non-zero current,
\begin{align}
j(x,t) &=\int_{-\infty}^{\infty} dk\,\frac{1}{\hbar} \frac{d E_k}{dk}\, \rho_W(x,k,t), \label{eq:j}\\
&=\int_{-\infty}^{\infty} dk\, \frac{1}{\hbar}\frac{d E_k}{dk}\, \rho_W^{(1)}(x,k,t). \nonumber
\end{align}
If we now substitute Eq.~(\ref{eq:expan}) into Eq.~(\ref{eq:masterWig}), disregard contributions of higher order in $\beta^{-1}$,  and separate symmetric and anti-symmetric  terms in  $k$, we  obtain
\begin{equation}
\frac{\partial n(x,t)}{\partial t}\,\rho_W^{(0)} (k)+\frac{1}{\hbar}\frac{\partial \rho_W^{(1)} (x,k,t)}{\partial x}\,\frac{d E_k}{dk}=0\;,\label{eq:even}
\end{equation}
\begin{equation}
\frac{\partial \rho_W^{(1)}(x,k,t)}{\partial t}+\frac{1}{\hbar}\frac{\partial n}{\partial x}\,\rho_W^{(0)} (k)\,\frac{dE_k}{dk}=-\tau_{s}^{-1}(k)\rho_W^{(1)}(x,k,t)\;,
\label{eq:odd}
\end{equation}
\noindent
where
\begin{equation}
\tau_{s}^{-1}(k)=\int_{-\infty}^{\infty}dq \, w_{k\rightarrow q}^{(0)}\;.\label{eq:ts}
\end{equation}
In the above equation we used the fact that  $\int dq \,w_{q\rightarrow k}^{(0)}\,\rho_W^{(1)}(q,x,t)=0$ due to (\ref{eq:sym}). This equation corresponds to a standard relaxation time approximation   in the transport theory.

Integrating (\ref{eq:even}) over $k$ and using (\ref{eq:j}),  we obtain the continuity equation
\begin{equation}
\label{eq:continuity}
\frac{\partial n(x,t)}{\partial t}+\frac{\partial j(x,t)}{\partial x}=0.
\end{equation}
Assuming that the relaxation time with respect to momentum is short compared to the time over which the density $n(x,t)$ evolves, we use the quasi-stationary solution of Eq.~(\ref{eq:odd}) for $\rho_W^{(1)}$,
\begin{equation}
\rho_W^{(1)}(x,k,t)=-\frac{\partial n(x,t)}{\partial x}\, \tau_s(k)\,\rho_W^{(0)}(k)\,\frac{1}{\hbar} \frac{dE_k}{dk}.\label{eq:P1}
\end{equation}
The current then is just a diffusive current, 
\begin{equation}
j(x,t)=-D\,\frac{\partial n(x,t)}{\partial x}\;,
\end{equation}
\begin{equation}
D=\int_{-\infty}^{\infty}dk\,\rho_W^{(0)} (k) \,\tau_s(k)\,\left(\frac{1}{\hbar}\frac{dE_k}{dk}\right)^2\;.
\label{eq:diff}
\end{equation}
\noindent
The continuity equation (\ref{eq:continuity}) takes the form of the diffusion equation for a spatial distribution $n(x,t)$,
\begin{equation}
\frac{\partial n(x,t)}{\partial t}=D\frac{\partial^2 n(x,t)}{\partial x^2}
\end{equation}
The explicit form of the diffusion coefficient  $D$ in the semiclassical region (\ref{eq:para}), which follows from Eqs.~(\ref{eq:wkq}),  (\ref{eq:ts}) and  (\ref{eq:diff}),  is given in Eq.~(10) of the main text.

\section{Renormalization of the fermion spectrum}
\label{sec:renormalization}

In addition to fermion scattering, coupling of the fermions to the bosonic field leads to a renormalization of the fermion energy spectrum (the polaronic effect), fermion mixing, and fermion-fermion interaction. For weak coupling, the corresponding effects are small.  It is the small renormalization condition that imposes a constraint on the coupling strength. We specify it here for the Ohmic-coupling, where the density of states of the bosonic bath weighted with the coupling is $2\hbar^{-2}\sum_\gamma \lambda_{\gamma n}^2\delta(\omega-\omega_\gamma) = \alpha\omega\exp(-\omega/\omega_c)$ for all lattice sites $n$. 

The effect of the Ohmic spin-boson coupling in an Ising chain is different from the case of a particle in a potential well coupled to bosons, where the coupling could be incorporated into the  potential \cite{Caldeira1983}. In the case of a spin chain,  the polaronic energy shift depends on the fermion energy and also on the transverse magnetic field.

Special attention has to be paid to the case of a very large parameter $\omega_c$. A simple perturbation theory shown below diverges if it is extended to bosons with energies $\hbar\omega_\gamma\to \infty$. However, it is clear on physical grounds that high-energy bosons with $\hbar\omega_\gamma\gg 2J$  should adiabatically follow the spin dynamics. For large $\omega_c$, we introduce a cutoff frequency $\omega_{\rm cutoff}$ such that $\omega_{\rm cutoff}\gg 2J/\hbar$ but $\omega_{\rm cutoff}  < \omega_c$. The effect of bosons with $\omega_\gamma\geq \omega_{\rm cutoff}$ can be accounted for by the standard polaronic transformation 
\[U=\exp\left[\sum_{\gamma,n}\sigma_n^x\frac{\lambda_{\gamma n}}{\hbar\omega_\gamma}(b_{\gamma n}- b^\dagger_{\gamma n})\Theta(\omega_\gamma-\omega_{\rm cutoff})\right]
\]
where $\Theta(x)$ is the step function. This transformation eliminates the coupling of $\sigma_n^x$ to such bosons. It shows that the major effect of the high-energy bosons is the renormalization of the Ising energy $J\to J\exp(-W)$ with $W\sim 2\alpha\log(\omega_c/\omega_{\rm cutoff})$. We assume that such renormalization has been done and that $W\ll 1$.

After the high-energy bosons are eliminated  (if they were present initially), the analysis of the renormalization of the fermion energy can be done using the explicit form of the parameters of the coupling Hamiltonian in  the main text,
\begin{align}
\label{eq:coupling_parameters}
&c_{k k'}=2N^{-1/2}\cos[(\theta_k+\theta_{k'})/2],\nonumber\\
 &s_{k k'}=iN^{-1/2}\sin[(\theta_k+\theta_{k'})/2].           
\end{align}
%
If we disregard the contribution of thermal bosons, to the second order in $\lambda_\gamma$  the expression for the polaronic energy shift $2J\Sigma_k$ of a fermion with wave vector $k$ has a standard form, with
\begin{align}
\label{eq:renormalization_single}
\Sigma_k=& \frac{\alpha}{2}\,{\rm v.p.}\int  \bar\omega \exp[-\bar\omega/\bar\omega_c]d\bar\omega\nonumber\\
&\times\sum_{k'}\left[\frac{|c_{kk'}|^2}{\epsilon_k-\epsilon_{k'} - \bar\omega} + \frac{4|s_{kk'}|^2}{\epsilon_k + \epsilon_{k'} - \bar\omega}\right],
\end{align}
where v.p. indicates the principal value of the integral and $\bar\omega_c = \hbar\omega_c/2J$. For $\omega_c\lesssim 2J/\hbar$, the integration over $\bar \omega$ goes from $\bar\omega=0$ to $\infty$. On the other hand, if $\omega_c\gg 2J/\hbar$, the upper limit of the integral is  $\bar\omega_{\rm cutoff} =  \hbar\omega_{\rm cutoff} /2J$. 

The coupling-induced mixing corresponds to an extra term in the fermion Hamiltonian of the form of $2J\sum_k\Sigma^{(c)}_k \eta_k^\dagger\eta_{-k}^\dagger +$~H.c.  If we disregard the contribution from thermally excited bosons,
\begin{align}
\label{eq:particle_hole}
\Sigma_k^{(c)}= & \frac{\alpha}{2}\,{\rm v.p.}\int  \bar\omega \exp[-\bar\omega/\bar\omega_c]d\bar\omega\sum_{k'}s_{kk'}c_{kk'}\nonumber\\
&\times\left[(\epsilon_k-\epsilon_{k'} - \bar\omega)^{-1} - (\epsilon_k + \epsilon_{k'} - \bar\omega)^{-1}\right].
\end{align}
The limits of the integral over $\bar\omega$ are the same as in Eq.~(\ref{eq:renormalization_single}).

It is important that the coupling to bosons does not lead to mixing of long-wavelength ($k\to 0$) excitations. This is because $\epsilon_{-k'}=\epsilon_{k'}$, whereas $s_{kk'}c_{kk'}\propto\sin(\theta_k+\theta_{k'})$ changes sign for $k'\to -k'$ in the limit $k\to 0$.

Of interest to us is the parameter range close to the critical point, $|g-1|\ll 1$, and a range of the scaled fermion energies $\epsilon_k\ll 1$. Because such fermions have small $k$, the coupling practically does not mix  fermions with opposite momenta. The leading-order scaled energy shift for $\epsilon_k \ll 1$ is $\Sigma_k\sim -\alpha\bar\omega_{\rm cutoff}$ for $\bar\omega_c\gg 1$, i.e., for broadband bosons. On the other hand, for narrow-band bosons (compared to the Ising coupling energy $J$), i.e., for $\bar\omega_c\ll 1$, we have $\Sigma_k\sim -\alpha\bar\omega_c^2$ for $k\to 0$. We note that the condition $\bar\omega_c\ll 1$ is compatible with the conditions $\bar\omega_c\gg 1/\beta, 1-g_{\rm opt}$ used in the main text to describe relaxation of long-wavelength fermions; here $g_{\rm opt}=1-x_{\rm opt}/\beta$, where $x_{\rm opt}$ is given by Eq.~(17) in  the main text; $1-g_{\rm opt} \ll 1$.

The shift $\Sigma_k$ for $k\to 0$ determines the shift in the critical value of the control parameter $g$. The shape of the spectrum of long-wavelength fermions near the critical point is not changed by the renormalization (\ref{eq:renormalization_single}). Indeed, it can be seen  from Eq.~(\ref{eq:renormalization_single}) that $\Sigma_k\approx \Sigma_{k\to 0} + C\epsilon_k$. Constant $C$ is $\sim \alpha\log\bar\omega_{\rm cutoff}$ for $\bar\omega_c\gg 1$ and is $\sim \alpha\bar \omega_c^2$ for $\bar\omega_c\ll 1$.

\section{\label{sec:QBE}  Quantum Boltzmann equation for fermion populations neglecting spatial fluctuations}

The goal of this section is to justify the equation that describes the evolution of the spatially-averaged fermion density due to generation-recombination processes, Eq.~(11) of the main text. For the scaled transverse field $g(t)> g_*$, i.e.,  prior to the crossover to a diffusion limited recombination, spatial fluctuations  of the fermion density can be disregarded [$g_*$ is given by Eq.~(15) of the main text].  For fermions with momentum $k$, time evolution of their  density  $\rho_k(t)\equiv \rho_{k,k}(t)=\langle \eta_{k}^{\dagger}(t) \eta_k(t)\rangle$   can be described by the quantum Boltzmann equation, cf. Ref.~\cite{Santoro:08},  which in standard notations has the form
\begin{equation}
\frac{\partial \rho_{k}}{\partial t}={\cal L}_{k}^{(a)}[\rho]+{\cal L}_{k}^{(b{})}[\rho],\label{eq:QBE}
\end{equation}
\noindent
\vspace{-0.2in}
\begin{equation}
{\cal L}_{k}^{(a)}[\rho] =\sum_{q}\left(W_{q k}^{+-}(1-\rho_k)\rho_q-W_{k q}^{+-} \rho_k(1-\rho_q)\right)\nonumber
\end{equation}
\noindent
\vspace{-0.2in}
\begin{equation}
{\cal L}_{k}^{(b{})}[\rho] = \sum_{q}\left(W_{k q}^{--}(1-\rho_k)(1-\rho_q)-W_{k q}^{++}\rho_k \rho_q)\right)\nonumber
\end{equation}
\noindent
Here   ${\cal L}^{(a)}$ describes inelastic intraband  scattering, 
( cf. Eq.~(8) in the main text). Operator ${\cal L}^{(b{})}$  describes two-fermion creation with rate $\propto W_{kq}^{--}$) and annihilation with rate $\propto W_{kq}^{++}$). The transition rates in   Born approximation are given in Eq.~(5) of the main text.
For fixed   $g$, Eq.~(\ref{eq:QBE}) has a stationary solution given by the Fermi-Dirac distribution with zero chemical potential,  $\rho_k=1/[\exp(2J \epsilon_k/k_B T)+1]$. In (\ref{eq:QBE}) we assumed that the inverse duration of a collision  is much smaller than the QA rate,  $k_B T/\hbar \gg |\dot g|$.

Unlike the scattering rates $W_{kq}^{+-}$, the  rates $W_{kq}^{\mu\mu}$ depend  exponentially strongly on the relation between the energy gap $ \Delta=2J|1-g|$ and $k_B T$.  At the initial stage of QA $\Delta\gg k_BT$ and the  system is mostly frozen in its ground state, because fermion generation is suppressed,  $W^{--}_{kq}\propto \exp(-2\Delta/k_B T)$.   As the critical region $\Delta \lesssim k_B T$ is traversed, fermions with energies $\lesssim k_B T$  become thermally excited (and are potentially also excited via the Kibble-Zurek mechanism, which in the considered case of small $|\dot g|$ gives less excitations).

After the critical point is passed,  the system again enters the semiclassical region $\Delta\gg k_BT$. The two-fermion generation rate  $W^{--}_{kq}$ slows down and  the  fermion population decreases. A key observation is that, since fermion annihilation  requires a two-fermion collision with rate  $W^{++}_{kq} \rho_k \rho_q$, it also slows down. In contrast, the rate of intraband scattering  described by the operator ${\cal L}_{k}^{(a)}$ in Eq.~(\ref{eq:QBE}) has  terms linear in  $\rho_k$, which do not contain exponentially small factors. Therefore intraband transitions are faster than interband transitions in the semiclassical region. 

The physical description of the dynamics is based on the idea that, because of the intraband scattering, there is first established thermal distribution within the fermion band. The total fermion population changes on a longer time scale due to  interband processes. If we  keep only the intraband scattering  terms in  Eq.~(\ref{eq:QBE}), this equation takes the form
\begin{equation}
\frac{\partial \rho_{k}}{\partial t} \simeq   \sum_{q} L^{(0)}_{k q} \rho_{q},\quad L^{(0)}_{k q}=W_{ q k}^{+-} - \delta_{k q } \sum_{k'}W_{k k'}^{+-} \label{eq:scat}
\end{equation}
\noindent
 In the limit of large $N$ for $|1-g|\ll 1$ we introduce  a scale-free integral kernel, 
 \begin{equation}
 L^{(0)}_{k q}=\tau_{r}^{-1}  \bar L^{(0)}_{KQ},\qquad K=k/k_{\rm th} 
 \end{equation}
where $k_{\rm th}$ is defined in (\ref{eq:K}) and 
\begin{align}
\bar L^{(0)}_{KQ}&= w_{QK} - \delta(K-Q) \int d K' \,\,w_{KK'}\;, \label{eq:PK} \\
w_{QK} &=\frac{1}{2}(Q^2-K^2)\{1-\exp[-(Q^2-K^2)/2]\}^{-1}.\nonumber
\end{align}
 \noindent
The relaxation rate $\tau_{r}^{-1}$ is given in Eq.~(9) of the main text. Except for the rescaling, the operator $\bar L^{(0)}$ has the same form as the operator $\hat {\cal L}^{(0)}$ in Eq.~(8) of the main text.

A direct calculation shows that the eigenvalues $e_m$ of $\bar L^{(0)}$ are non-positive. The maximal eigenvalue $e_0=0$  has as the eigenstate the stationary solution of (\ref{eq:scat}) 
 \begin{equation}
 \rho_k=\frac{\langle n \rangle}{n_{\rm th}} e^{-\beta \epsilon_k},\quad n_{\rm th}=\frac{1}{N}\sum_k e^{-\beta \epsilon_k}. \label{eq:G}
 \end{equation}
 \noindent
Equation (\ref{eq:G}) describes a  quasi-equilibrium thermal distribution over fermion momenta;  $\langle n \rangle $ is the spatially-averaged fermion density, and  $n_{\rm th}$ is the  thermal equilibrium density.

One further find from the analysis of the eigenvalues of the operator $\bar L^{(0)}$ that its eigenvalues   $e_{m>0}$ form a continuous spectrum (in the limit of $N\rightarrow \infty$)  with a gap given by the first nonzero eigenvalue $e_1=-6.6$. The rate $\tau_r^{-1}|e_1|$ is the  typical relaxation rate of fermion momenta. It increases with the distance $1-g\propto\Delta$ from the critical point. 
 
The density $\langle n\rangle$ varies on the time scale $t \gg \tau_r $. An equation describing the slow time evolution of $\langle n\rangle$  can be found by substituting  expression (\ref{eq:G}) into the full Boltzmann equation (\ref{eq:QBE}) and performing summation over the momentum $k$ in this equation. This gives Eq.~(11) of the main text.

\vspace{0.2in}

\section{\label{eq:cross} Crossover from the mean field regime to the diffusion limited  regime}

In this section we provide an alternative estimate of the fermion density where spatial fluctuations of the fermion density cannot be disregarded and the recombination rate becomes diffusion-limited. Such crossover has been studied in a number of papers \cite{Avraham1990,Privman1993,Lee:1994,Allam2013} (see also 
\cite{Tauber2014_1}) where the diffusion coefficient and the recombination rate were assumed constant. In the mean-field regime described by the rate equation, cf. Eq.~(11) of the main text, these assumptions lead to a linear increase of the reciprocal density in time,  if generation is neglected.  In our problem,  the  fermion density decreases logarithmically  slowly, Eq.~(14) of the main text,   while the diffusion rate $D\sim v_T^2\tau_r \propto (1-g)^{-3/2}$, Eq.~(10) of the main text, sharply  falls off  in  time.

To estimate the time and density where there occurs the  crossover in our problem, we consider a random spatial configuration of fermions  at an  instant $\tau$. Typically, a given  fermion is separated from other fermions on  the both sides by    \lq\lq empty" intervals \cite{Avraham1990} of size  $\ell (\tau) $= $\langle  n\rangle^{-1}(\tau) \ll \ell_{\rm mfp}(\tau)$.
For $t>\tau$,  the fermion  diffuses  toward the  boundaries of this interval, which are moving themselves due to fermion recombination. The recombination rate for the considered fermion at time $t$ is determined by the probability to have diffused over the distance $\ell(t)$. It has the form $w[2\pi \ell_D(t,\tau)]^{-1/2} \exp[-\ell^2(t)/2\ell_D^2(t,\tau)]$, where  $\ell_D(t,\tau)=[4 \int_{\tau}^{t} dt'    D(t')]^{1/2}$ is the diffusion length, $\ell_D^2(t,\tau)\propto [1-g(\tau)]^{1/2}-[1-g(t)]^{1/2}$.

 If the instant  $\tau$ corresponds to  a sufficiently large $g(\tau)<1$,  when  the diffusion  is fast,  then one can find such time $t$ that $\ell_D(t,\tau) \gg  \ell(t)$  and the considered fermion with high likelihood will   recombine with other fermions. However, for a later time $\tau$ this is not the case, because of the diffusion slowing down. In other words, the condition  $  \ell_D(t,\tau) > \ell(t)$ cannot be met.  The critical value of $\tau=t_*$ and the corresponding crossover density $n_*$ can be estimated from the condition   that  the curves $\ell_D(t,\tau)$ and $\ell(t)$ touch each other, $d (\ell_{D}^{2}(t,\tau)/dt = d\ell^2(t)/dt$.  If fermion generation can be neglected,  this gives
\begin{equation}
n(t_*)  = k w(t_*)/D(t_*), \nonumber
\end{equation}
\noindent
where $k\sim 1$. This is the condition given in Eq.~(15) of the main text. For $\tau > t_*$ spatial correlations in the fermion distribution become significant.


\begin{thebibliography}{99}
\bibitem{footnote1} The rate equation for $\langle n(t)\rangle$ (\ref{eq:generation_recombination}) can  be obtained also from the quantum   Boltzmann equation for  the occupation numbers $\langle \eta_{k}^{\dagger}(t) \eta_k(t)\rangle$, see Sec.~III of the SM. The Boltzmann equation applies also in the critical region \cite{Santoro:02}.
\end{thebibliography}
\end{document}